\documentclass[5p,sort&compress]{elsarticle}
\usepackage{amsmath}
\usepackage{amssymb}
\usepackage{graphicx}
\usepackage{comment}
\usepackage{dsfont}

\def\Slash#1{\setbox0=\hbox{$#1$} 
\dimen0=\wd0 
\setbox1=\hbox{/} \dimen1=\wd1 
\ifdim\dimen0>\dimen1 
\rlap{\hbox to \dimen0{\hfil/\hfil}} 
#1 
\else 
\rlap{\hbox to \dimen1{\hfil$#1$\hfil}} 
/ 
\fi}

\begin{document}

\begin{frontmatter}

\title{Delta-baryon mass in a covariant Faddeev approach}

\author{D.~Nicmorus, G.~Eichmann, A.~Krassnigg, R.~Alkofer}

\address{Institut f\"ur Physik, Karl-Franzens-Universit\"at Graz, A-8010 Graz, Austria }

\begin{abstract}
        We present a calculation of the three-quark core contribution to the mass of the $\Delta $-resonance 
        in a Poincar\'{e}-covariant Faddeev framework. A consistent setup for the dressed-quark propagator,
        the quark-quark and quark-'diquark' interactions is used, where all the ingredients are solutions of
        their respective Dyson-Schwinger or Bethe-Salpeter equations in rainbow-ladder truncation.
        We discuss the evolution of the $\Delta$ mass with the current-quark mass and compare to the previously
        obtained mass of the nucleon.
\end{abstract}


\end{frontmatter}

\section{Introduction}

        Experimentally, the first excited state of the nucleon, the $\Delta$
        resonance, is observed in scattering pions, photons, or electrons
        off nucleon targets. Recent precise experiments at LEGS, BATES, ELSA, MAMI, and JLAB
        do not only report on its mass ($M_\Delta=1.232$ GeV) and width ($\Gamma_\Delta = 120$ MeV)
        but also measure the electromagnetic $N\to \Delta \gamma$ transition form factors 
        \cite{Schmieden:2006xw,Sparveris:2004jn,Beck:1999ge,Blanpied:2001ae,Elsner:2005cz,Pospischil:2000ad,Stave:2006ea}.
        The basic properties of the lightest resonance of the nucleon have been calculated
        within various approaches such as constituent-quark models 
	\cite{Isgur:1979be,Buchmann:1996bd,Faessler:2006ky,Melde:2005hy,Melde:2007iu,Melde:2008yr,Melde:2008dg,Metsch:2003ix,Migura:2006en,Metsch:2008zz},
        Skyrme models \cite{Wirzba:1986sc,Abada:1995db,Walliser:1996ps},
        chiral cloudy bag models \cite{Kaelbermann:1983zb,Thomas:1982kv},
        chiral effective field theory methods \cite{Hemmert:1996xg,Hemmert:1997ye,Pascalutsa:2005ts,Hacker:2005fh,Pascalutsa:2005nd},
	and lattice-regularized QCD 
	\cite{Zanotti:2003fx,Gockeler:2007rx,WalkerLoud:2008bp,Gattringer:2008td,Alexandrou:2008bn}.

        Ultimately it is desirable to obtain a detailed understanding of the rich structure of $N$ and $\Delta$ 
	baryons and their properties in terms of QCD's quark and gluon degrees of freedom in a quantum-field 
	theoretical framework. The Dyson-Schwinger-equation (DSE) point of view employed in this letter offers 
	a non-perturbative continuum approach to QCD, reviewed recently in \cite{Fischer:2006ub,Roberts:2007jh}.
        This fully self-consistent infinite set of coupled integral equations provides a tool to access both 
	the perturbative and the non-perturbative regimes of QCD. The most prominent phenomena emerging in the 
	latter are dynamical chiral symmetry breaking and confinement, which, in the same way as bound states, require
	a nonperturbative treatment.
        
	Hadrons and their properties are studied in this approach via covariant bound-state 
	equations. While mesons can be described by solutions of the $q\bar{q}$-bound-state Bethe-Salpeter 
	equation (BSE), the case of a baryon is more involved and has therefore not yet been addressed at 
	the same level of sophistication. Based on the observation that the attractive nature of quark-antiquark 
	correlations in a color-singlet meson is also attractive for $\bar{3}_C$ quark-quark correlations 
	within a color-singlet baryon, a first step is to study a two-body problem by means of a quark-'diquark' 
	bound-state BSE. In a basic setup, parametrizations for the needed quark and diquark propagators and 
	diquark amplitudes were used to calculate $N$ and $\Delta$ masses \cite{Oettel:1998bk,Hecht:2002ej} and 
	nucleon electromagnetic form factors
        \cite{Oettel:1999gc,Oettel:2000jj,Alkofer:2004yf,Holl:2005zi,Cloet:2008wg}.  
        The next step on the way to the full covariant three-body bound-state equation is to replace the 
	parameterizations by solutions of the corresponding DS (quark propagator) and BS (diquark amplitudes) 
	equations in analogy to sophisticated meson studies. The simplest consistent setup of this kind is 
	the rainbow-ladder (RL) truncation, which was implemented in this fashion and discussed
	in full detail for the nucleon in \cite{Eichmann:2007nn}; for a short summary, see \cite{Eichmann:2008zz}.
	Recently, evidence has been provided  that in Landau gauge QCD all components of the
	quark-gluon vertex are infrared divergent \cite{Alkofer:2006gz,Alkofer:2008tt}. However, there are
	indications that for most hadronic observables the corresponding contributions beyond RL
	truncation are small. In this respect mesonic effects in the quark and quark-gluon-vertex DSEs, see e.g.\ 
	\cite{Pichowsky:1999mu,Fischer:2007ze,Fischer:2008wy}, are more important.
        Due to the existence of these and other contributions beyond RL truncation, 
	instead of aiming at actually reproducing light-quark baryon properties in RL truncation,
	a different approach was recently taken. It allows an identification of RL results with a
	hadron's 'quark core' via estimating all corrections that are expected beyond RL
	and subsequently tuning the RL contribution such that the corrected result would describe experimental data. 
	The corresponding prescription was introduced in \cite{Eichmann:2008ae} and applied
	to the nucleon mass and electromagnetic form factors in \cite{Eichmann:2008ef}.
        In the present work we adopt this procedure and calculate the respective core contribution to the mass of 
	the $\Delta (1232)$, a necessary step for a reliable analysis of the nontrivial $N\to \Delta \gamma$ transition.

        The paper is organized as follows: first we briefly summarize the Poincar\'{e}-covariant
        Faddeev approach to bary\-ons and its simplification to a quark-diquark picture. After discussing the 
	ingredients of the covariant quark-diquark BSE and the implications of the interaction as well
	as the diquark concept, we present the tensor decomposition of the $\Delta$ amplitude.
        Finally we discuss the results for the mass of the $\Delta$ baryon and compare to the
        mass of the nucleon obtained in the same approach, together with a selection of results from lattice QCD.

        Throughout this paper we work in Euclidean momentum space and use the isospin-symmetric limit $m_u=m_d$.

\section{Covariant Faddeev equations}

        Baryonic bound states correspond to poles in the three-quark scattering matrix, i.e. the amputated 
	and connected quark 6-point function. The residue at a pole associated to a baryon of mass $M$ defines 
	the respective bound-state amplitude. It satisfies a covariant homogeneous integral equation, which, upon
        neglecting irreducible three-body interactions, leads to a covariant equation of the Faddeev type \cite{Cahill:1988dx}.

        The binding of baryons in this framework is dominated by quark-quark correlations.
        In particular, a color-singlet baryon emerges as a bound state of a color-triplet quark and a
	color-antitriplet diquark correlation \cite{Cahill:1987qr}. The same mechanism that binds color-singlet 
	mesons is suitable to account for an attraction in the corresponding diquark channels
        \cite{Cahill:1987qr,Maris:2002yu}. To implement this concept, one approximates the quark-quark 
	scattering matrix that appears in the covariant Faddeev equations by a separable sum of 
	pseudoparticle-pole contributions corresponding to various diquark correlations. 
	This procedure leads to a quark-diquark BSE on the baryon's mass shell, where
        scalar $0^+$ and axial-vector $1^+$ diquarks have been used to describe 
	the nucleon. Of those only the axial-vector diquark correlation 
	contributes to the spin$-\frac{3}{2}$ and isospin$-\frac{3}{2}$ flavor symmetric $\Delta$;
	therefore its quark-diquark BSE reads \cite{Oettel:1998bk,Oettel:2000jj}
                   \begin{equation}\label{deltabse}
                        \Phi_{\alpha\beta}^{\mu\nu}(p,P) = \!\!\int\limits_k \!\! 
			\left\{ K^{\mu\rho}(p,k,P) \, S(p_q) \, \Phi^{\sigma\nu}(k,P) \right\}_{\alpha\beta}  D^{\rho\sigma}(p_d)\;,
                   \end{equation}
        where $P$ is the total baryon momentum, $p_q$, $p_d$ are quark and diquark momenta, $p$, $k$ are the 
	quark-diquark relative momenta, and $\int_k$ denotes $\int \frac{d^4 k}{(2\pi)^4}$.
        Greek superscripts represent Lorentz indices, greek subscripts fermion indices.
        The amplitudes $\Phi^{\mu\nu}(p,P)$ are the matrix-valued remainders of the full
        quark-diquark amplitude $\Phi_{\alpha\beta}^{\mu\nu}(p,P)\,u^\nu_\beta(P)$ for the $\Delta$, where $u^\nu_\beta(P)$ is a Rarita-Schwinger spinor
        describing a free spin-3/2 particle with momentum $P$.
        \begin{figure}[htp]
                    \begin{center}
                    \includegraphics[scale=0.26]{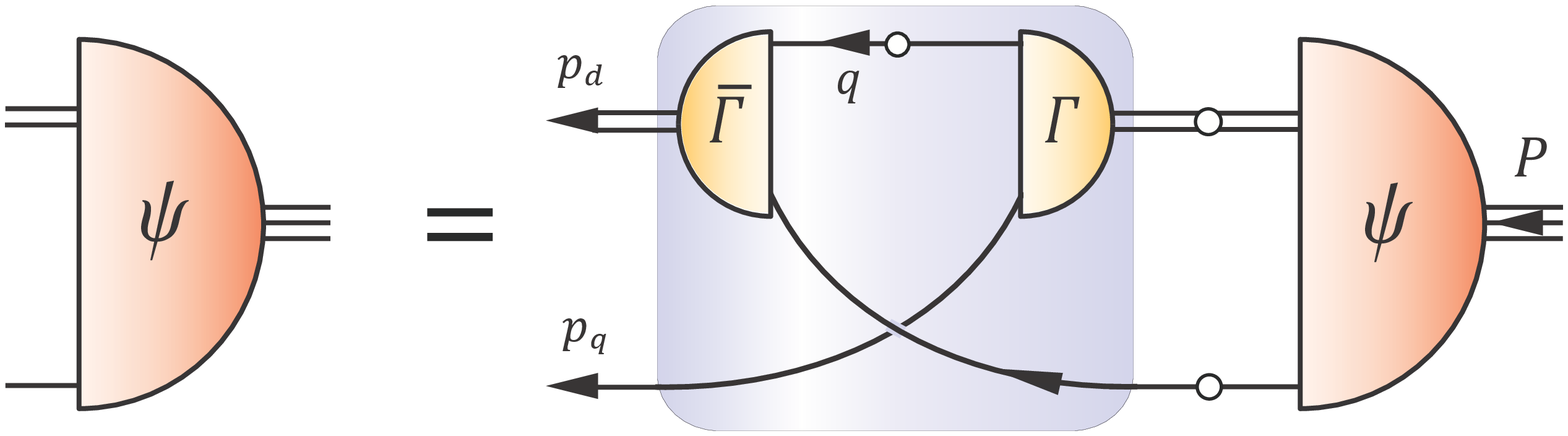}
                    \caption{The quark-diquark BSE, Eq.\,(\ref{deltabse})}\label{fig1}
                    \end{center}
        \label{fig.1}
        \end{figure}
        In addition one needs to specify the dressed-quark propagator $S$, the diquark propagator $D^{\rho\sigma}$, 
	and the axial-vector diquark amplitude $\Gamma^\nu$ and its charge-conjugate $\bar{\Gamma}^\nu$ which appear 
	in the quark-diquark kernel:
            \begin{equation}\label{deltabse-kernel}
                 K^{\mu\nu}_{\alpha\beta}(p,k,P) = \left\{ \Gamma^\nu(k_r,k_d) \, S^T(q) \, \bar{\Gamma}^\mu(p_r,-p_d) \right\}_{\alpha\beta},
            \end{equation}
        where subscripts $"r"$ denote quark-quark relative momenta and $"d"$ diquark momenta.
        By virtue of Eqs.~(\ref{deltabse}) and (\ref{deltabse-kernel}) the mechanism which binds the $\Delta$ is 
	an iterated exchange of roles between the single quark and any of the quarks contained in the diquark.
        This exchange is depicted in the quark-diquark BSE in Fig.~\ref{fig1}. We proceed by detailing the 
	ingredients of Eq.~(\ref{deltabse}).

\section{Quark propagator and effective coupling}\label{sec:interaction}

        We begin with the renormalized dressed-quark propagator $S(p)$. It is expressed in terms of 
	two scalar functions,
            \begin{equation}\label{solqprop}
                S^{-1}(p) = A(p^2)\,\left( i\Slash{p} + M(p^2) \right),
            \end{equation}
        where $1/A(p^2)$ is the quark wave-function renormalization and $M(p^2)$ the 
	renormalization-point independent quark mass function. The quark propagator satisfies 
	the quark DSE (also referred to as the QCD gap equation):
            \begin{equation}\label{quarkdse}
                S^{-1}_{\alpha\beta}(p) = Z_2 \left( i\Slash{p} + m_\textrm{bare} \right)_{\alpha\beta}  - 
		\int\limits^\Lambda_q \mathcal{K}_{\alpha\gamma,\delta\beta}(p,q) \,S_{\gamma\delta}(q),
            \end{equation}
        where $Z_{2}$ is the quark renormalization constant and $\Lambda$ a regularization mass-scale.
        The bare current-quark mass $m_\textrm{bare}$ serves as an input of the equation and is related
        to the renormalization-independent current mass $\hat{m}$ via one-loop evolution:
        $m_\textrm{bare} = \hat{m}/[\ln{ (\Lambda/\Lambda_{QCD})}]^{\gamma_m}$,
        where the anomalous dimension of the quark mass function is $\gamma_m=12/(11N_C-2N_f)$.
        We use $N_f=4$ and $\Lambda_{QCD}=0.234$ GeV.
        In the chiral limit, $\hat{m}=0$.

        The kernel $\mathcal{K}$ of the quark DSE includes the dressed gluon propagator as well as one bare 
	and one dressed quark-gluon vertex. The fully dressed gluon propagator and quark-gluon vertex could 
	in principle be obtained as solutions of the infinite coupled tower of QCD's DSEs together with all 
	other Green functions of the theory, cf.~\cite{Alkofer:2008tt} and references therein. 
	In practical numerical studies one employs a truncation of the
	infinite system of equations by solving only a subset explicitly. Green functions appearing in the 
	subset but not solved for are represented by substantiated ans\"atze.
        In connection with the simultaneous solution of a meson BSE, it is imperative to employ a truncation 
	that preserves the axial-vector
        Ward-Takahashi identity if one wants to correctly implement chiral symmetry and its dynamical breaking.
        This ensures a massless pion in the chiral limit --- the Goldstone boson related to dynamical chiral 
	symmetry breaking --- and leads to a generalized Gell-Mann--Oakes--Renner relation for all pseudoscalar
	mesons and all current-quark masses \cite{Maris:1997hd,Maris:1997tm,Holl:2004fr}.
        The lowest order in such a symmetry-preserving truncation scheme \cite{Munczek:1994zz,Bender:1996bb},
        the RL truncation, has been extensively used in DSE studies of mesons, 
	e.\,g.~\cite{Alkofer:2002bp,Maris:2006ea} (spectroscopy), 
        \cite{Holl:2005vu,Maris:2005tt,Bhagwat:2006pu} (electromagnetic properties), and references therein.

        Out of the 12 general covariants of the quark-gluon vertex, the RL truncation retains only its vector 
	part $\sim \gamma^\mu$. The non-perturbative dressing of the gluon propagator and the quark-gluon 
	vertex are absorbed into an effective coupling $\alpha(k^2)$ which is given by an ansatz invoked by 
	the truncation. The kernel $\mathcal{K}$ of the quark DSE then reads:
            \begin{equation}\label{RLkernel}
                \mathcal{K}_{\alpha\gamma,\beta\delta}(p,q) =  Z_2^2 \, \frac{\scriptstyle 4\pi \alpha(k^2)}{\scriptstyle k^2} \, T^{\mu\nu}_k
                                                     \!\left( i\gamma^\mu\frac{\scriptstyle \lambda^a}{\scriptstyle 2}\right)_{\alpha\gamma}
                                                     \!\left( i\gamma^\nu\frac{\scriptstyle \lambda^a}{\scriptstyle 2}\right)_{\beta\delta},
            \end{equation}
        where $\lambda^a$ are the $SU(3)_C$ Gell-Mann matrices. $T^{\mu\nu}_k=\delta^{\mu\nu} - \hat{k}^\mu \hat{k}^\nu$
        is a transverse projector with respect to the gluon momentum $k=p-q$, and $\hat{k}^\mu=k^\mu/\sqrt{k^2}$ 
	denotes a normalized 4-vector. A convenient form of $\alpha(k^2)$ is given by 
	\cite{Maris:1999nt}
                    \begin{equation}\label{couplingMT}
                        \alpha(k^2) = \frac{c\pi}{\omega^7} \,\left(\frac{k^2}{\Lambda_0^2}\right)^2 \!\!
                                      e^{\frac{-k^2}{\omega^2\Lambda_0^2}} + \frac{\scriptstyle \pi\gamma_m\,\left( 1-e^{-k^2/\Lambda_0^2} \right) }
                                      {\ln\sqrt{\scriptstyle e^2 -1 + \left( 1+k^2/\Lambda_{QCD}^2 \right)^2 } }
                    \end{equation}
        with a scale $\Lambda_0=1$ GeV.
        The second term reproduces QCD's perturbative running coupling and decreases logarithmically for large gluon-momenta.
        The first term of the interaction accounts for the non-perturbative enhancement at small and intermediate gluon 
	momenta: it provides the necessary strength to allow for dynamical chiral symmetry breaking and the dynamical 
	generation of a constituent-quark mass scale. With the UV part fixed by perturbative QCD, the coupling is 
	characterized by two parameters, the strength $c$ and width $\omega$ of the interaction in the infrared.
        The interaction of Eq.~(\ref{couplingMT}) provides a reasonable description of masses, decay constants 
	and electromagnetic properties of ground-state pseudoscalar and vector-mesons with equal-mass constituents 
	up to bottomonium if the coupling strength is kept fixed for all values of the quark mass (see \cite{Maris:2006ea} 
	and references therein). In these investigations a value $c= 0.37$ has commonly been used, which reproduces 
	the phenomenological quark condensate and the experimental decay constant $f_\pi=131$ MeV at a
	pion mass of $m_\pi=140$ MeV.
        Furthermore, pseudoscalar- and vector-meson masses and decay constants have turned out to be insensitive upon 
	variation of the coupling-width parameter $\omega$ in the range $\omega \approx 0.4 \pm 0.1$ \cite{Maris:1999nt}.

        Naturally, any description making use of a truncation is a priori incomplete. On the one hand,
	corrections come from pseudoscalar meson-cloud contributions which were introduced by the cloudy-bag model 
	\cite{Theberge:1980ye,Thomas:1982kv,Lu:1997sd} and systematically 
	incorporated into chiral perturbation theory \cite{Gasser:1983yg,Bernard:1995dp}.
        These corrections provide a substantial attractive contribution to the 'quark core' of dynamically 
	generated hadron observables in the chiral regime, whereas they vanish with increasing current-quark mass.
        Their impact on the chiral structure of the quark mass function and condensate, $f_\pi$, $m_\rho$, 
	and nucleon and $\Delta$ observables has been demonstrated in the NJL-model \cite{Oertel:2000jp,Cloet:2008fw}, 
	DSE studies \cite{Pichowsky:1999mu,Hecht:2002ej,Fischer:2007ze,Fischer:2008wy}, and chiral 
	extrapolations of lattice results \cite{Young:2002cj}. On the other hand, a resummation of non-resonant 
	diagrams beyond RL provides further attraction in the pseudoscalar and vector-meson 
	channels \cite{Bhagwat:2004hn,Matevosyan:2006bk}. The above corrections are in agreement with the 
	quark-gluon-vertex DSE and the infrared properties of its solution \cite{Alkofer:2008tt}.
	In order to anticipate such corrections, a mere RL result 
	should systematically overestimate the experimental masses. More concretely, in 
	\cite{Eichmann:2008ae,Eichmann:2008ef} a parameterization for the relation $m_\rho(m_\pi)$ was used
	to characterize the quark-core contribution to the meson masses:
            \begin{equation}\label{core:mrho}
                x_\rho^2 = 1 + x_\pi^4/(0.6+x_\pi^2), \;\; x_\rho = m_\rho/m_\rho^0, \;\; x_\pi = m_\pi/m_\rho^0,
            \end{equation}
        with the chiral-limit value $m_\rho^0 = 0.99$ GeV. Starting from this value, the sum of chiral corrections would
        reduce $m_\rho$ in the chiral limit by $\sim 25\%$ whereas above the $s$-quark mass the quark core approaches 
	corresponding lattice results for $m_\rho$.
        In order to reproduce Eq.~(\ref{core:mrho}) by solving the $\rho$-meson BSE, the coupling strength $c$
        of Eq.~(\ref{couplingMT}) has to depend on the current-quark mass
	thereby reflecting the properties discussed above. 
	The following fit yields Eq.~(\ref{core:mrho}):
            \begin{equation}\label{core:C}
                 c(\omega, \hat{m}) = 0.11 +  \frac{ 0.86 \,b(\Delta\omega) }{ 1 + 0.885\,x_q + (0.474\,x_q)^2}\;, 
            \end{equation}
        where $x_q=\hat{m}/(0.12\textrm{GeV})$. At each value of the current-quark mass $\hat{m}$, the parame\-terization
            \begin{equation}\label{core:B}
                 b(\Delta\omega) =  1 - 0.15\,\Delta\omega + (1.50\,\Delta\omega)^2 +(2.95\,\Delta\omega)^3
            \end{equation}
        fixes the result for $m_\rho$ to the same value in the range $\omega = \bar{\omega}(\hat{m}) \pm |\Delta\omega|$,
        with the central value given by $\bar{\omega}(\hat{m}) = 0.38 + 0.17/(1+x_q)$.
        As demonstrated in Refs.~\cite{Eichmann:2008ae,Eichmann:2008ef}, this procedure induces consistent (overestimated) 
	values for a range of $\pi$, $\rho$ and nucleon observables as obtained from their respective meson 
	and quark-diquark BSEs. These results are approximately $\omega$-independent
        for $|\Delta\omega| \lesssim 0.1$, i.e. in the region where $b(\Delta\omega) \approx 1$.

        In the present work we employ this 'core model' of Eqs.~(\ref{couplingMT}) - (\ref{core:B}) to directly compare
        the obtained $\Delta$ mass with the previously obtained result for the nucleon. We note here that all parameters
	of the interaction were fixed using information from the $\pi$ and $\rho$ masses.

\section{Diquarks}
	It follows from the separable diquark-pole ansatz for the two-quark scattering matrix that 
        the axial-vector diquark amplitude in Eq.~(\ref{deltabse-kernel}) satisfies a diquark BSE
        with the on-shell condition $P^2=-M_\textrm{av}^2$ for the total diquark momentum $P^\nu$, which reads
                    \begin{equation}\label{diqbse}
                        \Gamma^\mu_{\alpha\beta}(p,P) =\int\limits^\Lambda_q 
                        \mathcal{K}_{\alpha\gamma,\beta\delta}(p,q) \left\{ S(q_+) \Gamma^\mu(q,P) S^T(q_-) \right\}_{\gamma\delta},
                    \end{equation}
        where $p$ is the relative momentum between the two quarks in the diquark bound state
        and $q_\pm = \pm q + P/2$ are the quark momenta. Greek subscripts represent Dirac and color 
	indices of the involved quarks. The form of the corresponding scalar-diquark BSE is identical 
	to Eq.~(\ref{diqbse}) if the Lorentz superscript $\mu$ is dropped.

        Consistency with the meson sector implies the identification of the irreducible two-quark 
	kernel $\mathcal{K}$ in the diquark BSE with the RL-truncated kernel of Eq.~(\ref{RLkernel}).
        This setup (self-consistently) yields real diquark masses. While this feature does not explicitly 
	contradict diquark confinement (see e.g. \cite{Alkofer:2000wg,Roberts:2000aa}), we note here that
        diquark bound-state poles are likely to be removed from the real timelike $P^2$ axis by large repulsive corrections
        beyond RL truncation \cite{Bender:1996bb,Hellstern:1997nv,Bender:2002as,Bhagwat:2004hn}.
        Assuming that also a two-quark scattering matrix free of real timelike poles contains scales characterizing the 
	diquark correlations, it is nonetheless meaningful to make use of the diquark concept. 
	Evidence in this direction also comes from Coulomb-gauge QCD, where diquark correlations
	have recently been shown to retain their size while being removed from the physical
	spectrum \cite{Alkofer:2005ug}.

        The axial-vector diquark amplitude can be decomposed into a sum of $12$ Lorentz-invariant dressing functions
	(a hat on a four-momentum indicates a normalized vector),
                    \begin{equation}
                        \Gamma^\mu_{\alpha\beta}(q,P)  =
                        \sum_{k=1}^{12} f_k^\textrm{av}(q^2,\hat{q}\!\cdot\!\hat{P},P^2) \left\{ i\tau^\mu_k(q,P)\,C \right\}_{\alpha\beta},
                    \end{equation}
        of which, due to transversality at the axial-vector diquark pole, only $8$ contribute on-shell.
        A suitable orthonormal basis for this on-shell part of the axial-vector amplitude is given 
	by (cf.~App.~B.2 of Ref.~\cite{Eichmann:2007nn}):
                    \begin{align}\label{axbasis}
                       \nonumber  \tau_1^{\mu} &= \gamma^{\mu}                                            &   \tau_5^{\mu} &= \hat{q}\!\cdot\!\hat{P}\,( \gamma^{\mu} \Slash{\hat{q}}_T - \hat{q}^{\mu} )                            \\
                       \nonumber  \tau_2^{\mu} &= \gamma^{\mu}\Slash{\hat{P}}                             &   \tau_6^{\mu} &= \frac{\scriptstyle i \gamma^{\mu}}{\scriptstyle 2} [ \Slash{\hat{P}},\Slash{\hat{q}} ] + i\hat{q}^{\mu}\Slash{\hat{P}}         \\
                                  \tau_3^{\mu} &= i\,\hat{q}^{\mu}                                          &   \tau_7^{\mu} &= \hat{q}^{\mu}\Slash{\hat{q}}_T - \frac{\scriptstyle \hat{q}^2_T}{\scriptstyle 3}\gamma^{\mu}                                \\
                       \nonumber  \tau_4^{\mu} &= \hat{q}\!\cdot\!\hat{P}\,\hat{q}^{\mu} \Slash{\hat{P}}  &   \tau_8^{\mu} &= \frac{\scriptstyle \hat{q}^{\mu}}{\scriptstyle 2} [ \Slash{\hat{P}},\Slash{\hat{q}} ] + \frac{\scriptstyle \hat{q}^2_T}{\scriptstyle 3}\gamma^{\mu} \Slash{\hat{P}} 
                    \end{align}
	 where the subscript $T$ denotes transverse projection.
	 
        The diquark amplitude is a solution of Eq.~(\ref{diqbse}) only on the diquark's mass shell.
        However, within a baryon the two-quark system moves at all possible values of the total diquark momentum.
        Therefore, to fully obtain the kernel of the quark-diquark BSE (\ref{deltabse}), one has to 
	specify the off-shell behavior of the diquark amplitude and propagator as well.
        For completeness we recapitulate the off-shell extension introduced in Ref.~\cite{Eichmann:2008ef}:
        we use an on-shell approximation for the dressing functions: $f_k^\textrm{av}(q^2,\hat{q}\!\cdot\!
	\hat{P},P^2) \approx f_k^\textrm{av}(q^2,0,-M_\textrm{av}^2)$.
        Assuming that only the dominant amplitude remains relevant at large $P^2$,
        each subleading amplitude $f_{k>1}^\textrm{av}$ is suppressed by a factor $g(x) = 1/(x+2)$, 
	where $x=P^2/M_\textrm{av}^2$; and each occurrence of $\hat{P}$ in the basis (\ref{axbasis}) 
	is augmented by a factor $h(x) = -i\sqrt{x/(x+2)}$ to ensure a sensible analytic continuation 
	for off-shell momenta which does not alter the power-law behavior at $P^2\rightarrow\infty$. 
        The off-shell ans\"atze $g(x)$ and $h(x)$ leave the BSE solutions on the diquark's mass shell 
	invariant: $g(-1) = h(-1) = 1$.

        Reinsertion of the diquark pole ansatz into the Dyson series for the two-quark T-matrix yields 
	the following expression for the diquark propagator \cite{Oettel:2002wf,Eichmann:2007nn}:
            \begin{align}
                D^{-1}_{\mu\nu}(P) &= M_\textrm{av}^2 \left\{ \lambda \,\delta_{\mu\nu} +
                                   \beta\,F_{\mu\nu}(x) + Q_{\mu\nu}(x) \right\}, \label{dqprop}\\
                Q^{\mu\nu}(x) &= \frac{1}{2M_\textrm{av}^2}\! \int\limits_q \!
		\textrm{Tr}_D\! \left\{ \bar{\Gamma}^\mu(q,-P) S(q_+) \Gamma^\nu(q,P) S^T(q_-)\right\}
            \end{align}
        with $\lambda = -Q_T(-1)$ and $\beta = 1-Q_T'(-1)$. 
	$Q_T = Q_{\mu\nu}Q_{\mu\nu}/3$ denotes the transverse
        contribution to the quark-loop integral $Q_{\mu\nu}$. The two-loop 
	integral $F_{\mu\nu}(x)$ emerges via inclusion of the subleading amplitudes $f_{k>1}^\textrm{av}$.
        The ansatz
            \begin{equation}\label{dqprop:F}
                F_{\mu\nu}(x) = \delta_{\mu\nu} \left( 1-1/(x+2)^2 \right)/2
            \end{equation}
        ensures the on-shell behavior $D^{-1}_T(P^2 \rightarrow - M_\textrm{av}^2) \longrightarrow P^2 + M_\textrm{av}^2$
        and a satisfactory approximation in the UV extracted from the result of the numerical two-loop calculation.
        We note that the systematic uncertainty connected to the diquark amplitude's off-shell ansatz
        is minimized through use of the consistent expression for the diquark propagator, Eq.~(\ref{dqprop}).
        While a different choice for the off-shell functions $g(x)$ and $h(x)$ changes the values of 
	$F(x)$ and $Q(x)$ accordingly and necessitates a modified ansatz (\ref{dqprop:F}),
        it has no material impact on calculated observables since only the product of two diquark amplitudes and 
	the propagator (i.\,e., the quark-quark T matrix) enters the covariant Faddeev equation's kernel.

\section{Quark-diquark amplitudes for the $\Delta$}

        \begin{figure*}
                    \begin{center}
                    \includegraphics[scale=0.83]{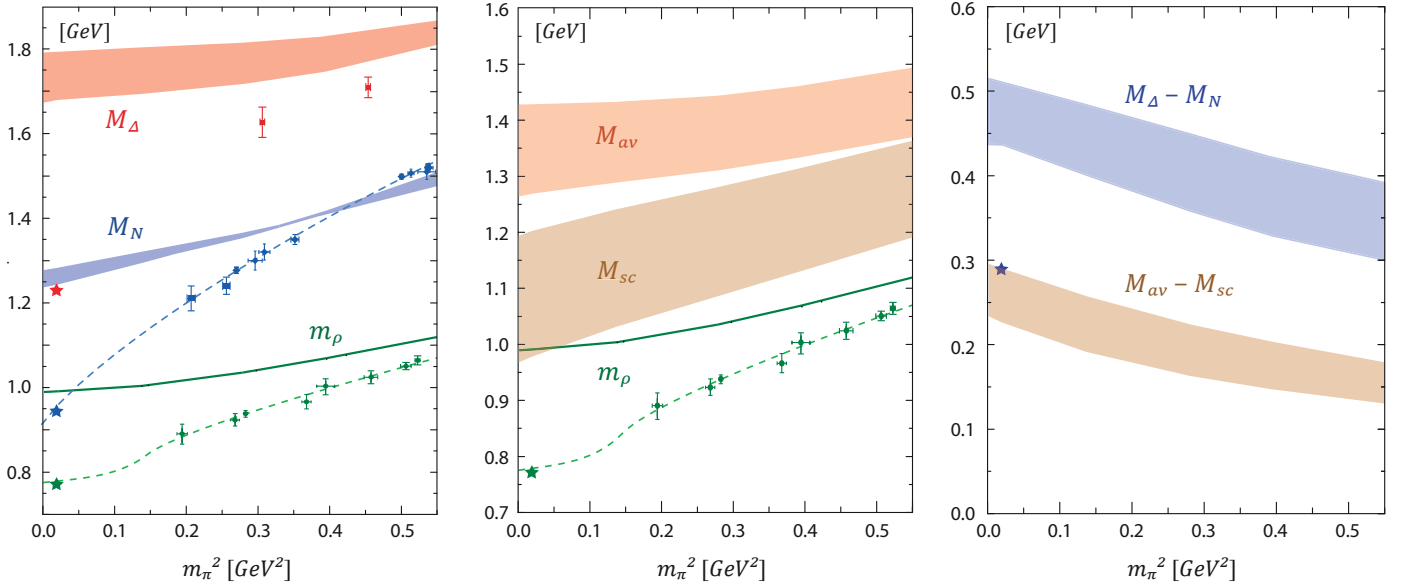}
                    \caption{Evolution of $N$ and $\Delta$ masses (\textit{left panel}), 
		             evolution of $m_\rho$, scalar and axial-vector diquark masses (\textit{center panel}),
                             and the mass splittings $M_\Delta-M_N$ and $M_\textrm{av}-M_\textrm{sc}$ (\textit{right panel})
                             vs.~pion mass squared.
                             The bands denote the sensitivity to a variation of $\omega = \bar{\omega}(\hat{m}) \pm 0.07$.
                             The solid curve for $m_\rho$ represents the chosen input of Eq.\,(\ref{core:mrho});
                             nucleon and diquark masses were calculated in Ref.~\cite{Eichmann:2008ef}.
                             We compare to a selection of lattice data and their chiral extrapolations (\textit{dashed lines}) for
                             $m_\rho$ \cite{AliKhan:2001tx,Allton:2005fb}, $M_N$ \cite{AliKhan:2003cu,Frigori:2007wa,Leinweber:2003dg}
                             and $M_\Delta$ \cite{Zanotti:2003fx}.
                             Stars denote the experimental values.\label{fig:2}} 
                    \end{center}
        \end{figure*}

        Finally, to compute the amplitude and mass of the $\Delta$ baryon numerically,
        the structure of the on-shell quark-diquark amplitude $\Phi$ of Eq.\,(\ref{deltabse}) needs to be specified.
        Now denoting the total $\Delta$ momentum by $P$, with $P^2=-M_\Delta^2$ and $\hat{P} = P/(iM_\Delta)$,
        it is decomposed into 8 covariant structures \cite{Oettel:1998bk}:
                    \begin{equation}\label{deltaamplitdecompos}
                        \Phi_{\alpha\beta}^{\mu\nu}(p,P)\! =\!\! \sum_{k=1}^8 f_k^\Delta(p^2,\hat{p}\cdot\hat{P}) \Big\{\!   \tau_k^{\mu \rho}(p,P)\,\Lambda_+(P)\,\Lambda^{\rho \nu}(P) \!\Big\}_{\alpha\beta}\,,
                    \end{equation}
        where the basis elements include the Rarita-Schwinger projector onto positive-energy and spin-$3/2$ spinors, 
	construc\-ted from $\Lambda_+(P) = (\mathds{1}+\Slash{\hat{P}})/2$ and
                    \begin{equation}
                        \Lambda^{\mu \nu}(P) = \delta^{\mu\nu} + \frac{1}{3}\left( \hat{P}^\mu \gamma^\nu - \hat{P}^\nu \gamma^\mu - \gamma^\mu \gamma^\nu - 2 \hat{P}^\mu \hat{P}^\nu \right).
                    \end{equation}
        The number of basis elements can be inferred from the Clifford algebra.
        A general Green function with two fermion legs, two vector legs and two independent momenta 
	$p$ and $P$ allows for 40 possible Dirac covariants $\tau_{k=1\dots 40}^{\mu\rho}$, where four come from 
	$\delta^{\mu\rho} \left\{ \mathds{1}, \,\Slash{p}, \,\Slash{P}, \,\Slash{p}\,\Slash{P} \right\}$ and 36 from
	all possible combinations of the $3\times 3\times 4$ terms
        $\left\{ p^\mu, P^\mu, \gamma^\mu \right\}$ $\times$ $\left\{ p^\rho, P^\rho, \gamma^\rho \right\} $ $\times$ 
	$\left\{ \mathds{1}, \,\Slash{p}, \,\Slash{P}, \,\Slash{p}\,\Slash{P} \right\}$.
        The elements $P^\rho$, $\gamma^\rho$, $\Slash{P}$ and $\Slash{p}\,\Slash{P}$ become redundant upon contraction 
	with the Rarita-Schwinger projector:
        $P^\rho \Lambda_+\,\Lambda^{\rho \nu} = \gamma^\rho \Lambda_+\,\Lambda^{\rho \nu} = 0$,
        $\left\{ \Slash{P}, \,\Slash{p}\,\Slash{P} \right\} \Lambda_+ = \left\{ \mathds{1}, \,\Slash{p} \right\} \Lambda_+$.
        This leaves 8 Dirac covariants for which a convenient orthogonal set is given by:
                    \begin{align}\label{deltaavbasis}
                        \tau_1^{\mu\rho} & =  \delta^{\mu\rho}     
                     &   \tau_2^{\mu\rho} & =  \frac{\scriptstyle 1}{\scriptstyle \sqrt{5}}\left(2\gamma_T^\mu  \,q^\rho 
        		-3 \delta^{\mu\rho}\Slash{q}\right)  \nonumber \\
                        \tau_3^{\mu\rho} & =  -\sqrt{3}  \,\hat{P}^\mu q^\rho \Slash{q}   
                     &   \tau_4^{\mu\rho} & =  \sqrt{3} \,\hat{P}^\mu q^\rho   \\
                        \tau_5^{\mu\rho} & =  - \gamma_T^\mu  \,q^\rho \Slash{q}   
	             &   \tau_6^{\mu\rho} & =  \gamma_T^\mu \,q^\rho \Slash{q} -\delta^{ \mu \rho}  -  3 \, q^\mu q^\rho   \nonumber   \\
                        \tau_7^{\mu\rho} & =  -\gamma_T^\mu \,q^\rho     
                     &   \tau_8^{\mu\rho} & =  \frac{\scriptstyle 1}{\scriptstyle \sqrt{5}}\!\left( \delta^{\mu \rho}\Slash{q} + \gamma_T^\mu \,q^\rho +   5\,  \,q^\mu q^\rho \Slash{q}  \right) \,, \nonumber
                    \end{align}
        with $q^\mu := i \,T^{\mu\nu}_P \hat{p}^\nu/\sqrt{1-(\hat{p}\!\cdot\!\hat{P})^2}$ and 
	$T^{\mu\nu}_P = \delta^{\mu\nu}-\hat{P}^\mu\hat{P}^\nu$.
        The corresponding orthogonality relation reads
                    \begin{equation}
                        \textrm{Tr}\left\{ \bar{\sigma}_k^{\mu\rho}(p,-P) \,\sigma_l^{\mu\rho}(p,P) \right\} = 4 \,\delta_{kl} (-1)^{k+1},
                    \end{equation}
        where $\sigma_k^{\mu \nu} (p,P) = \tau_k^{\mu \rho}(p,P)\,\Lambda_+(P)\,\Lambda^{\rho \nu}(P)$ and the conjugated amplitude 
        is defined as (the superscript ${}^T$ denotes the Dirac transpose):
                    \begin{equation}
                        \bar{\sigma}_k^{\mu\rho}(p,-P) = C\,\sigma_k^{\mu\rho}(-p,-P)^T C^T.
                    \end{equation}
        The basis (\ref{deltaavbasis}) corresponds to a partial-wave decomposition in terms of eigenfunctions of the
        quark-diquark total spin and orbital angular momentum in the $\Delta$ rest frame \cite{Oettel:1998bk,Oettel:2000ig}.
        Since there is only one spherically symmetric $s$-wave component ($f_1^\Delta$),
        the $\Delta$'s deviation from sphericity $=1$ can be explained by an admixture of
        $p$ ($f_{2,4,7}^\Delta$), $d$ ($f_{3,5,6}^\Delta$) and $f$ ($f_8^\Delta$) waves,
        which contribute a significant amount of orbital angular momentum to its amplitude.

        The isospin matrices of the $\Delta$ quark-diquark amplitude are constructed via removal of the diquark contributions
        from the full $\Delta$ flavor-amplitude that is obtained by the Clebsch-Gordan prescription:
            \begin{align}
                 &\Delta^{++} = \left(  u , 0 , 0 \right),   
                 &\Delta^{+}  = \left(  \frac{1}{\sqrt{3}} \,d , \frac{2}{\sqrt{3}} \,u , 0 \right), 
		 \nonumber \\
                 &\Delta^{0}  = \left( 0 , \frac{2}{\sqrt{3}} \,d , \frac{1}{\sqrt{3}} \,u \right),   
                 &\Delta^{-}  = \left( 0 , 0 , d \right)\;.
            \end{align}
        The first, second, and third entries in each vector correspond to the diquark's three symmetric 
	isospin-1 states. 
        Computing the flavor traces of the quark-diquark BSE leads to a global factor $1$ in the equal-quark-mass case.
        The color matrix $\delta_{AB}/\sqrt{3}$
        attached to each quark-diquark amplitude entails a global color factor $-1$ in the BSE.

        The standard procedure to solve the quark-diquark BSE (\ref{deltabse}) is detailed in Ref.\,\cite{Oettel:2001kd} for the analogous case of a nucleon.
        It involves a Chebyshev expansion in the angular variable $\hat{p}\!\cdot\!\hat{P}$ and
        leads to coupled one-dimensional eigenvalue equations for the Chebyshev moments of the dressing functions $f_k^\Delta$.
        They match the BSE solution at $P^2=-M_\Delta^2$, i.\,e.~for an eigenvalue $\lambda(P^2)=1$.

\section{Results and discussion}

        We solved Eqs.~(\ref{deltabse}), (\ref{quarkdse}), and (\ref{diqbse}) numerically using all ingredients as specified above and
	show the results in Fig.~\ref{fig:2}. The left panel depicts our calculated 'quark core' values for
	$m_\rho$, $M_N$ and $M_\Delta$, each together with a selection of lattice results and their chiral 
	extrapolations (if available). As abscissa values we use the corresponding $m_\pi^2$, in our calculation 
	obtained from the pseudoscalar $q\bar{q}$ BSE. The solid curve for $m_\rho$ is the input defined in
        Eqs.~(\ref{couplingMT}) - (\ref{core:B}), which completely fixes the parameters in our interaction.
        The bands represent our results for $M_N$ and $M_\Delta$ and explicitly show the sensitivity on the width 
	parameter $\omega$ as described in Sec.~\ref{sec:interaction}.
        At the physical pion mass we obtain $M_\Delta = 1.73(5)$ GeV; the corresponding mass for the nucleon 
	$M_N=1.26(2)$ GeV was reported in Ref.~\cite{Eichmann:2008ef}. At larger quark masses the deviation from 
	the lattice data diminishes. This result is in accordance with the assumption of Eq.~(\ref{core:mrho}),
        namely that beyond-RL corrections to hadronic observables (except cases like highly excited states
	or low-$x$ physics) become negligible in the limit of heavy quarks.
        
        It is instructive to compare our results to core masses estimated via, 
        e.\,g., the cloudy-bag model \cite{Pearce:1986du}, NJL model \cite{Ishii:1998tw}, 
	nucleon-pion Dyson-Schwinger studies \cite{Hecht:2002ej,Oettel:2002cw}, and
        a chiral analysis of lattice results \cite{Young:2002cj,Young:2002ib}.
	These predictions lie in the range $\sim 200-400$ MeV for the correction to 
	the nucleon mass induced by pseudoscalar meson loops; pseudoscalar-meson contributions 
	to $M_\Delta$ are expected to be of a similar size or even smaller than those to the nucleon.
	In this respect, our value for $M_N$ in the chiral region is roughly consistent with a 
	pseudoscalar-meson dressing providing the dominant correction to the quark-diquark 
	core, whereas the result for $M_\Delta$ is not, which indicates a larger correction.
        As a concrete example, consider the reduction of $N$ and $\Delta$ masses by 
	intermediate $N\pi$ and $\Delta\pi$ states, estimated in the chiral limit by the 
	expressions (e.\,g., \cite{Young:2002cj})
        \begin{align}
            M_N      &= M_N^\textrm{core}       + \Sigma_N, \quad       
            M_\Delta = M_\Delta ^\textrm{core} + \Sigma_\Delta, \nonumber \\
        \Sigma_N      &= - \lambda \int\limits_0^\infty dx \, x^2 \, u(x)^2 \left( 1 + \frac{32}{25} \,\frac{x}{x+1} \right), \label{chpt1}\\
	\Sigma_\Delta &= - \lambda \int\limits_0^\infty dx \, x^2 \, u(x)^2 \left( 1 + \frac{ 8}{25} \,\frac{x}{x-1} \right). \label{chpt2}
	\end{align}
        $k=x\,\Delta M$ is the pion momentum, $\Delta M=0.29$ GeV the physical 
	$N$--$\Delta$ mass splitting, and $\lambda = 3g_A^2 (\Delta M)^3/(8\pi^2 f_\pi^2)$ 
	with $g_A=1.26$ and $f_\pi=131$ MeV. The $NN\pi$, $\Delta\Delta\pi$ and $N\Delta\pi$ vertex 
	dressings $u(x)$ were assumed identical. Choosing a dipole form factor 
	$u(x) = \left\{ 1 + (k/\Lambda)^2 \right\}^{-2}$ with a regulator $\Lambda=0.8$ GeV 
	yields $\Sigma_N = -0.32$ GeV and $\Sigma_\Delta=-0.28$ GeV. Together with the experimental 
	numbers for $M_N$ and $M_\Delta$, these values provide the simple estimates $M_N^\textrm{core} \sim
	1.25$ GeV and $M_\Delta^\textrm{core} \sim 1.5$ GeV.

            \begin{table*}
                \begin{center}

                \begin{tabular}{l||c|c||c|c|c|c|c||c|c||c|c}
                                & $c$  &  $\omega$ & $\Lambda_\textrm{IR}$  & $\langle\bar{q}q\rangle^{1/3}_{1\,\textrm{GeV}}$
                                                                                                     &   $m_\rho$     &  $f_\pi$      & 
												     $f_\rho$     &  $M_\textrm{sc}$  & 
$M_\textrm{av}$  &  $M_N$        &  $M_\Delta$     \\ \hline\hline
                 Phen./Exp.         &                &           &                      & $0.236$        &   $0.77$       &  $0.131$      &  $0.216$      &                 &                 &  $0.94$       &  $1.23$        \\ \hline
                 Set A          & $0.37$         &  $0.40$   &   $0.72$             & $0.235$        &   $0.73$       &  $0.131$      &  $0.208$      &  $0.81$         &  $1.00$         &  $0.94$       &  $1.28$        \\
                 Set B (Core)   & $0.93$         &  $0.54$   &   $0.98$             & $0.319$        &   $0.99$       &  $0.176$      &  $0.280$      &  $1.08$         &  $1.35$         &  $1.25$       &  $1.72$        \\  \hline
                 (Set A)/(Set B)&                &           &   $0.73$             & $0.74$         &   $0.74$       &  $0.74$       &  $0.74$       &  $0.75$         &  $0.74$         &  $0.75$       &  $0.74$
                \end{tabular}
                 \caption{Comparison of several quantities calculated
                          from the quark DSE ($\langle\bar{q}q\rangle$), 
                          meson BSE ($m_\rho$, $f_\pi$, $f_\rho$), diquark BSE ($M_\textrm{sc}$, $M_\textrm{av}$),
                          and quark-diquark BSE ($M_N$,$M_\Delta$). 
                          For the explicit calculation of the meson, diquark and nucleon observables we refer 
			  the reader to Refs.~\cite{Eichmann:2007nn,Eichmann:2008ae,Eichmann:2008ef}.
                          The results correspond to a current mass $\hat{m}=6.1$ MeV which is related to the 
			  physical pion mass $m_\pi=138$ MeV.
                          The first row quotes experimental or phenomenological values.
                          Sets A and B are distinguished by different values for the input parameters
			  $c=(\Lambda_\textrm{IR}/\Lambda_0)^3$ and $\omega$;
                          set B represents the inflated quark core according to Eqs.~(\ref{core:C}) - (\ref{core:B}). 
                          Only the central $\omega$ values are shown (note that these need not be identical 
			  to the $\omega$-band averages quoted in the text.)
                          The units of the first three rows (except $c$ and $\omega$) are GeV.
                          The last row plots the ratios of sets A and B.}  \label{tab:1}

                \end{center}
            \end{table*}

        In this context one has to keep in mind that the identification the baryonic 
	quark core of Eqs.~(\ref{chpt1}) - (\ref{chpt2}) with the quark-diquark 'core' is more 
	complicated than in the meson case. More precisely, Eq.~(\ref{core:mrho}) assumes that corrections 
	to $m_\rho$ are partly induced by a pseudoscalar-meson cloud, and to a lesser extent related to 
	non-resonant corrections to RL truncation. In the baryon one has additional lines of improvement:
	inserting further diquark channels, abandoning the diquark pole ansatz in favor of the full $qq$ scattering
	kernel, and including irreducible 3-body interactions could affect $N$ and $\Delta$ properties differently, 
	but still describe a quark core in the sense of Eq.~(\ref{core:mrho}).
        
        We also note that the solution for $M_\Delta$ exhibits a sizeable $\omega$ dependence,
        a feature not present in $\pi$, $\rho$ and nucleon observables \cite{Eichmann:2008ae,Eichmann:2008ef}. 
        The scalar and axial-vector diquark masses exhibit particularly large sensitivities 
	to $\omega$ (see center panel of Fig.~\ref{fig:2}) which apparently cancel upon constituting the 
	nucleon mass. In the $M_\Delta$ case, the same consideration could suggest that taking into account only 
	an axial-vector diquark may not be sufficient for describing the $\Delta$, 
        and that a possible further isospin-$1$ (tensor) diquark component with a mass large enough to be 
	irrelevant for the nucleon could diminish the $\Delta$ 'core' mass.

        A further remark concerns the coupling strength $c$ of Eq.~(\ref{couplingMT}) at or close to the 
	chiral limit. Specifically, a comparison of the 'core' value induced by Eq.~(\ref{core:C}) and the 
	input used in Ref.~\cite{Maris:1999nt} provides valuable insight.
	The aim of \cite{Maris:1999nt} was to reproduce $\pi$ and $\rho$ properties;
        in addition, in our present setup the model also yields $N$ and $\Delta$ masses that are close to the
        experimental values ('set A' in Table~\ref{tab:1}; the 'core' values are summarized as 'set B'). 
	While this result is perhaps incidental since a 
	RL description is not likely to provide the complete underlying physical picture, it is still remarkable 
	that the 'set A'-to-'set B' ratios of all observables considered here are essentially identical. This is shown 
	in Table~\ref{tab:1} at the $u/d$-quark mass and can be understood as follows.
        Introducing $\eta=c^{1/3}/\omega$ and the scale $\Lambda_\textrm{IR}=c^{1/3} \Lambda_0$, the
        infrared contribution to the effective coupling $\alpha(k^2)$ can be rewritten as
            \begin{equation}
                \alpha_\textrm{IR}(k^2) = \pi \,\eta^7 x^2 e^{-\eta^2 x}, \quad x = k^2/\Lambda_\textrm{IR}^2.
            \end{equation}
        An insensitivity of certain observables under variation of the width $\omega$ at a certain coupling 
	strength $c$ translates into an invariance with respect to $\eta$.
        Furthermore, the combined increase of $c$ and $\omega$ according to Eqs.~(\ref{core:C}) - 
	(\ref{core:B}) to arrive at the 'quark core' changes $\eta$ by $\lesssim 5\%$;
        hence it can be viewed as a rescaling \textit{only} of $\Lambda_\textrm{IR}$.
        This quantity defines the only relevant scale in the chiral-limit RL quark DSE and meson/diquark BSEs
        where no interference with a finite current-quark mass is possible.
        If the renormalization point is chosen large enough, a rescaling of $\Lambda_\textrm{IR}$ equally affects
        the chiral-limit values of mass-dimensionful observables; and it induces scale invariance of dimensionless quantities.
        The effect on dynamically generated observables is still approximately valid at the small $u/d$ current-quark mass.
        It makes clear that any RL model which is able to reproduce experimental results for a given set of observables
        will, upon entering its 'core version', necessarily overestimate those results by the \textit{same} percentage.

        Finally, in the right panel of Fig.~\ref{fig:2} we plot the $N$--$\Delta$ mass splitting.
        According to Eqs.~(\ref{chpt1}) - (\ref{chpt2}), the pseudoscalar meson contribution to 
        the experimental value $M_\Delta-M_N=0.29$ GeV is small and positive: 
	$\Sigma_\Delta-\Sigma_N=0.04$ GeV, cf.~Ref.~\cite{Young:2002cj}.
        This is apparently not the case in our calculation, where at the $u/d$ mass 
	$(M_\Delta-M_N)^\textrm{core}=0.48(4)$ GeV and therefore predicts a negative correction to 
	the full splitting. We also compare $M_\Delta-M_N$ with the diquark mass splitting 
	$M_\textrm{av}-M_\textrm{sc}$. Both decrease with increasing current-quark mass; 
        nevertheless there is no direct relationship between the two quantities,
        since the axial-vector diquark contribution to the mass of the nucleon does not vanish.

\section{Conclusions}

        We calculated the mass of the $\Delta$-baryon in a covariant Faddeev approach, where
        the $\Delta$ quark core is pictured as a quark-diquark bound state. 
        The kernel of the quark-diquark Bethe-Salpeter equation is fully specified by the solutions of
        the Dyson-Schwinger equation for the dressed-quark propagator and the diquark Bethe-Salpeter equation 
	which are both solved in rainbow-ladder truncation.
        The physical input is specified by an effective interaction $\alpha(k^2)$ that is constructed to
        describe an overestimated quark core for the $\rho$ meson such that the (overall attractive) corrections 
	to rainbow-ladder truncation, 
	as e.\,g., a pseudoscalar meson-cloud dressing, are anticipated in the chiral regime.
        The calculation does not rely on any baryonic input, i.\,e., all parameters in the interaction are fixed
	via $m_\rho(m_\pi)$.
        Our result for the evolution of $M_\Delta$ with $m_\pi^2$ is justified a posteriori: at the $u/d$ 
	current-quark mass, $M_\Delta=1.73(5)$ GeV, whereas at larger quark masses our curve approaches 
	lattice data for $M_\Delta$. A relatively pronounced interaction dependence as well as a significant
	overestimation of the expected $M_\Delta$ core mass indicate that further diquark degrees of 
	freedom besides the axial-vector diquark could play a role in the construction of the $\Delta$ quark core.
        The approach presented herein can be systematically improved by eliminating the diquark ansatz in favor of 
	a sophisticated 3-quark interaction kernel and by developing suitable methods to directly incorporate 
	meson-cloud effects.

\section{Acknowledgements}

        We would like to thank J.~Berges, C.\,S.~Fischer, and C.\,D.~Roberts for helpful discussions.

        This work has been supported by the Austrian Science Fund FWF under Projects No.~P20592-N16, P20496-N16,
	and Doctoral Program No.~W1203 as well as by the BMBF grant 06DA267.

\bigskip

\bibliographystyle{elsart-num-mod}
\bibliography{had_nucl_graz}

\end{document}